\documentclass[10pt]{elsarticle}
\usepackage[utf8]{inputenc}
\usepackage[margin=3cm]{geometry}
\usepackage{amsmath}
\usepackage{graphicx}
\usepackage[labelfont=bf]{caption}
\usepackage[title]{appendix}
\usepackage{wrapfig}
\usepackage{multicol}
\usepackage{floatrow}
\usepackage{xcolor}
\usepackage{siunitx}
\usepackage{gensymb}
\usepackage[parfill]{parskip}
\usepackage{textcomp}
\usepackage{mathtools}
\usepackage{color}
\usepackage{longtable}
\let\vec\mathbf
\usepackage{indentfirst}
\usepackage{booktabs}
\usepackage{tabularx}
\newcommand{\subsectionbf}[1]{\subsection{\textbf{#1}}}
\newcolumntype{P}[1]{>{\centering\arraybackslash}p{#1}}
\usepackage[breaklinks=true,colorlinks=true,linkcolor=blue,urlcolor=blue,citecolor=blue]{hyperref}
\biboptions{sort&compress}
\usepackage{cleveref}
\usepackage{physics}
 
\newcommand{\blue}[1]{{\color{blue}#1}} 

\makeatletter
\def\ps@pprintTitle{%
  \let\@oddhead\@empty
  \let\@evenhead\@empty
  \def\@oddfoot{\reset@font\hfil\thepage\hfil}
  \let\@evenfoot\@oddfoot
}
\makeatother
\usepackage{setspace}
\begin{document}

\begin{frontmatter}
% ==================
% TITLE AND AUTHORS
% ==================

\title{GSPD: An algorithm for time-dependent tokamak equilibria design}    
\author[Princeton]{J.T. Wai}
\ead{jwai@princeton.edu}
\author[Princeton,PPPL]{E. Kolemen}
\ead{ekolemen@princeton.edu}
\address[Princeton]{Princeton University, Princeton, New Jersey, USA}
\address[PPPL]{Princeton Plasma Physics Laboratory, Princeton, New Jersey, USA}

% ===========
% ABSTRACT
% ===========
\begin{abstract}

\indent 
One of the common tasks required for designing new plasma pulses or shaping scenarios is to design the desired equilibria using an equilibrium (Grad-Shafranov equation) solver. However, standard equilibrium solvers are time-independent and cannot include dynamic effects such as plasma current drive, induced vessel currents, or voltage constraints. In this work we present the Grad-Shafranov Pulse Design (GSPD) algorithm, which solves for sequences of equilibria while simultaneously including time-dependent effects. The computed equilibria satisfy both Grad-Shafranov force balance and axisymmetric conductor circuit dynamics. The code for GSPD is available at \blue{github.com/plasmacontrol/GSPD}).

\end{abstract}

\end{frontmatter}

% ======================
% SECTION: INTRODUCTION
% ======================
\section{Introduction and example} \label{sec:Intro}

The Grad-Shafranov Pulse Design (GSPD) algorithm is a time-dependent equilibrium solver capable of optimizing for sequences of equilibria while including dynamic effects. The governing equations for this system are the Grad-Shafranov force balance condition,

\begin{equation}\label{eq:GS}
    \begin{split}
        \Delta^* \psi &= -\mu_0 r J_\phi \\
        J_\phi &=  rP'(\psi) + \frac{FF'(\psi)}{\mu_0 r}
    \end{split}    
\end{equation}

axisymmetric conductor dynamics, 

\begin{equation}\label{eq:circuit}
    v_i = R_i I_i + M_{ij} \dot I_j + \dot \Phi_i^{pla} 
\end{equation}

and plasma current evolution,

\begin{equation}\label{eq:psibry_dynamics}
    V_p = -\dot \psi_{bry} = R_p I_p + \frac{1}{I_p}  \dv{}{t} \left(\frac{1}{2}L_I I_p^2 \right).
\end{equation}

where,

\begin{equation*}
\begin{split}
     \Delta^*(\cdot) &= r\frac{\partial }{\partial r} \left( \frac{1}{r} \frac{ \partial (\cdot) }{\partial r } \right) + \frac{\partial^2 (\cdot) }{\partial z^2} = \text{Grad-Shafranov operator} \\
     \mu_0 &= \text{vacuum permeability constant} \\
     r &= \text{radial coordinate} \\
     J_\phi &= \text{toroidal current density distribution} \\
     P &= \text{pressure} \\
     \psi &= \text{poloidal flux per unit radian} \\
     F &= rB_t = \text{radius times toroidal magnetic field}. \\
     v &= \text{applied power supply voltage} \\
     i,j &= \text{indices for conducting elements} \\
     R_i &= \text{resistance in conductor i} \\
     I_i &= \text{current in conductor i} \\
     M_{ij} &= \text{mutual inductance between conductors i and j} \\
     \dot \Phi_i^{pla} &= \text{plasma coupling term, discussed in text} \\
     V_p &= \text{plasma loop voltage} \\
     \psi_{bry} &= \text{flux at plasma boundary} \\
     R_p &= \text{total plasma resistance} \\
     I_p &= \text{total plasma current} \\
     t &= \text{time} \\
     L_I &= \text{plasma internal inductance (unnormalized)}
\end{split}
\end{equation*}

\Cref{eq:GS} describes force balance in a tokamak and is used in standard equilibrium solvers. \Cref{eq:circuit} describes the circuit dynamics for all the axisymmetric conducting elements. The currents in the shaping coils and vacuum vessel structures evolve according to the applied voltage, their resistance, and induced flux via the mutual inductance. The $\dot \Phi_i^{pla}$ term is a plasma-coupling term which refers to induced flux due to plasma sources. That is, if the current in a conductor changes, this exerts a force on the plasma which then shifts to satisfy force balance. The shifting in the plasma current distribution also induces flux at all the other conductors. \Cref{eq:psibry_dynamics} describes the plasma current evolution due to surface voltage induced at the plasma boundary. The form chosen here is based on the Poynting flux method derived in \cite{Ejima1982}, although other forms are sometimes chosen in the shape control literature. 

Note that these are roughly the same governing equations that are used in the shape control literature, which are used in the development of control algorithms and plasma flight simulators (e.g. \cite{Welander2019}). However, GSPD is unique from plasma flight simulators in that it is an equilibrium design tool, not a control simulator. While they both include time-dependent effects, flight simulators require fully-developed feedback controllers including vertical stability control, and requires an equilibrium linearization step (which introduces modelling assumptions) that GSPD avoids.

\begin{figure}[H] 
    \begin{center}
        \makebox[\textwidth][c]{\includegraphics[width=5cm]{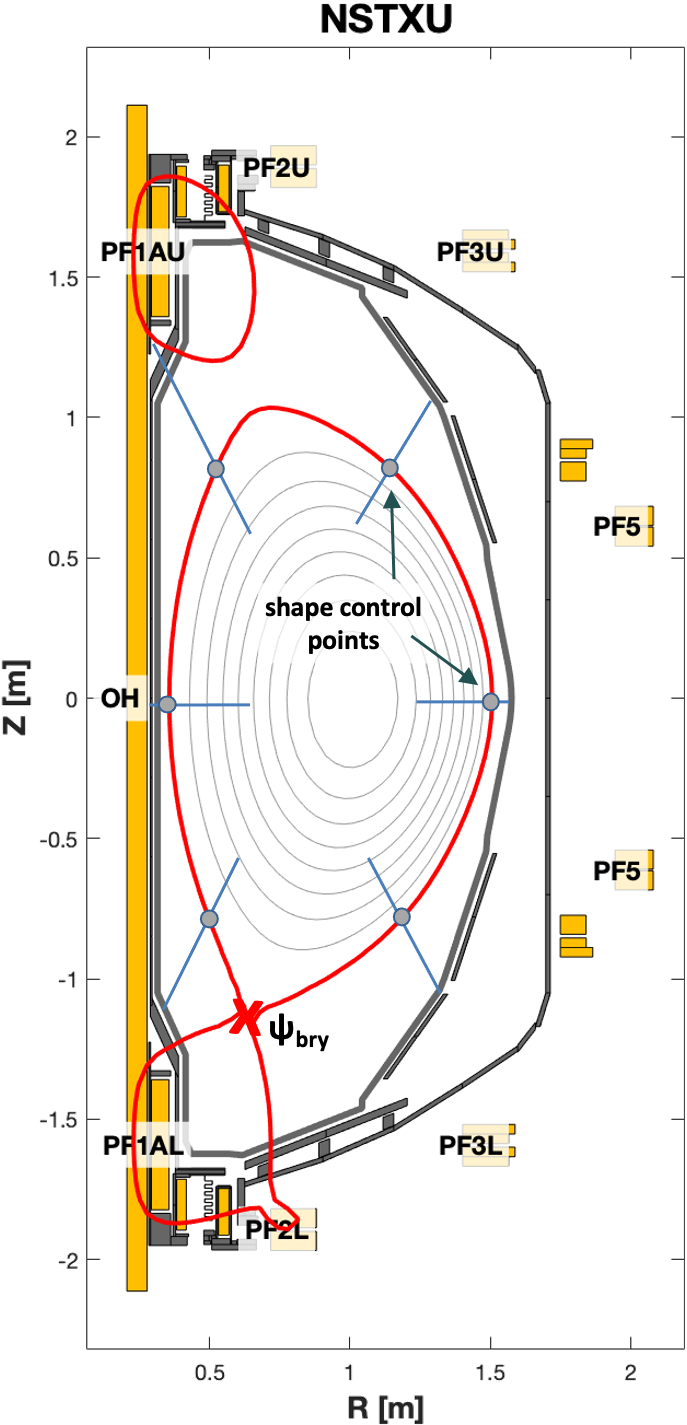}}
    \end{center}
    \caption{NSTX-U equilibrium geometry and control points defining a target shape.}
    \label{fig:geo_nstxu}
\end{figure}

Using GSPD requires that the user specify a series of target shapes to achieve specified by points on the plasma, for example, as shown in \cref{fig:geo_nstxu}. Then, GSPD optimizes the following cost function. This cost function has quadratic penalties on the control inputs (power supply voltages) and shaping errors. The weighting matrices are used to tune the equilibria design. Details are explained more rigorously in the next section. 

\begin{equation}
    J = \sum_{k=1}^{N} u_k^T \vec W_1 u_k + \Delta u_k^T \vec W_2 \Delta u_k + e_{k+1}^T \vec W_3 e_{k+1} + \Delta e_{k+1}^T \vec W_4  \Delta e_{k+1}
\end{equation}

GSPD alternates between two steps, updating the plasma current distribution to satisfy the Grad-Shafranov equation, and optimizing the current trajectories to achieve the target shapes. This is a special arrangement of the problem structure for satisfying the governing equations (\cref{eq:GS,eq:circuit,eq:psibry_dynamics}). While there is no standard method for feedforward design, most previous strategies would rely on solving for a particular equilibrium, linearizing the model, then performing some sort of optimization to step towards the next time point and equilibrium. It is difficult to design optimal trajectories with this type of strategy since it only looks ahead one step at a time. It can also be very computationally heavy, as optimization iterations may require for the same equilibrium to be solved for more than once. 

\begin{figure}[H] 
    \begin{center}
        \makebox[\textwidth][c]{\includegraphics[width=10cm]{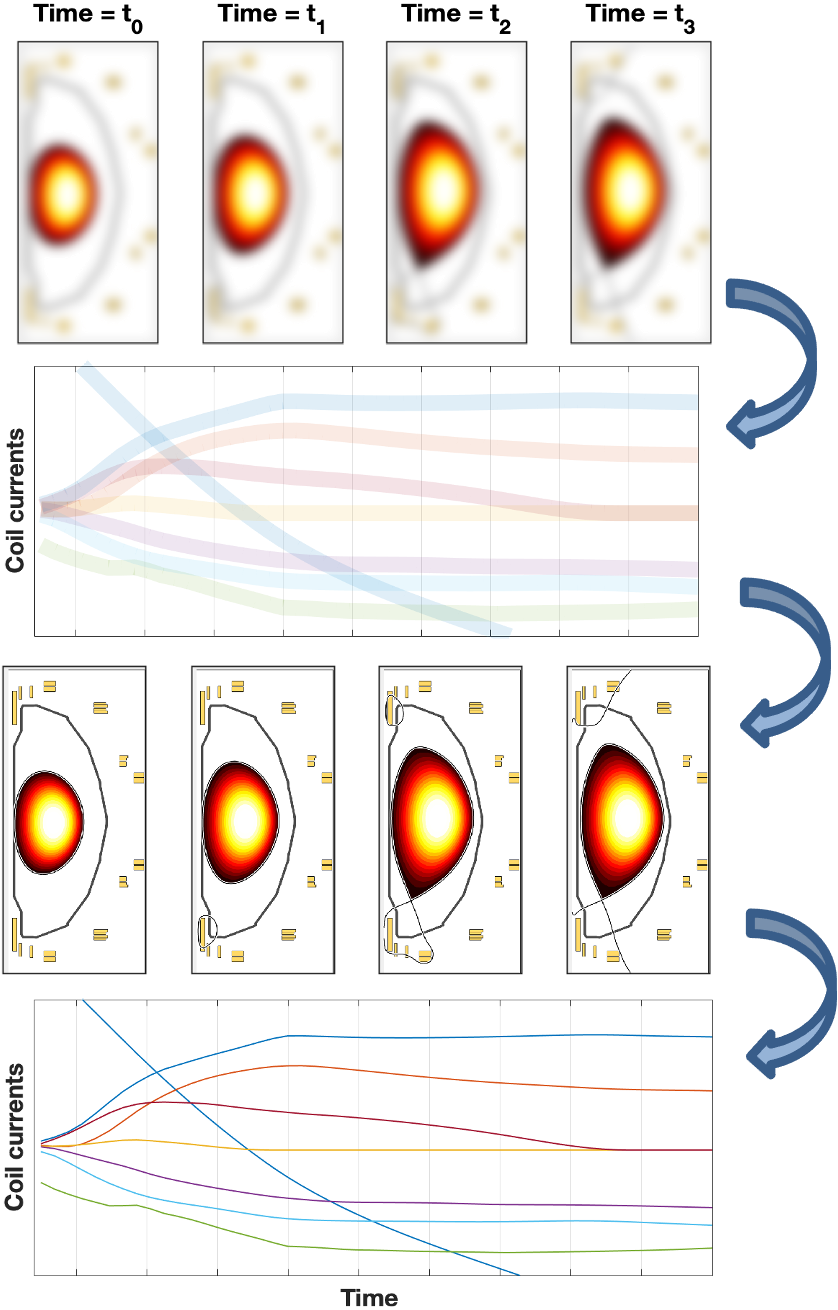}}
    \end{center}
    \caption{Schematic of the algorithm process. The algorithm begins with a rough, un-converged estimate of the entire sequence of equilibria. Then, an optimization step plans the current evolution, before refining the estimate of the Grad-Shafranov flux distribution. The algorithm alternates between these two steps. A traditional method for feedforward design would be to solve for a single equilbirum timeslice and achieve complete convergence, then linearize the model, step forward in time, and solve for subsequent equilibrium. The GSPD method here allows for easier time-dependent penalties, such as having smooth trajectories.}
    \label{fig:gspd_alg}
\end{figure}

A cartoon of this problem structure is given in \cref{fig:gspd_alg}. In GSPD, we converge towards an entire sequence of equilibria simultaneously. During the currents-update step, we solve an optimization problem for the entire trajectory. This allows for applying penalties to features of the dynamics, such as the smoothness of the trajectories. Then in the plasma-update the current distributions are updated to satisfy the GS force balance. 

\begin{figure}[H] 
    \begin{center}
        \makebox[\textwidth][c]{\includegraphics[width=14cm]{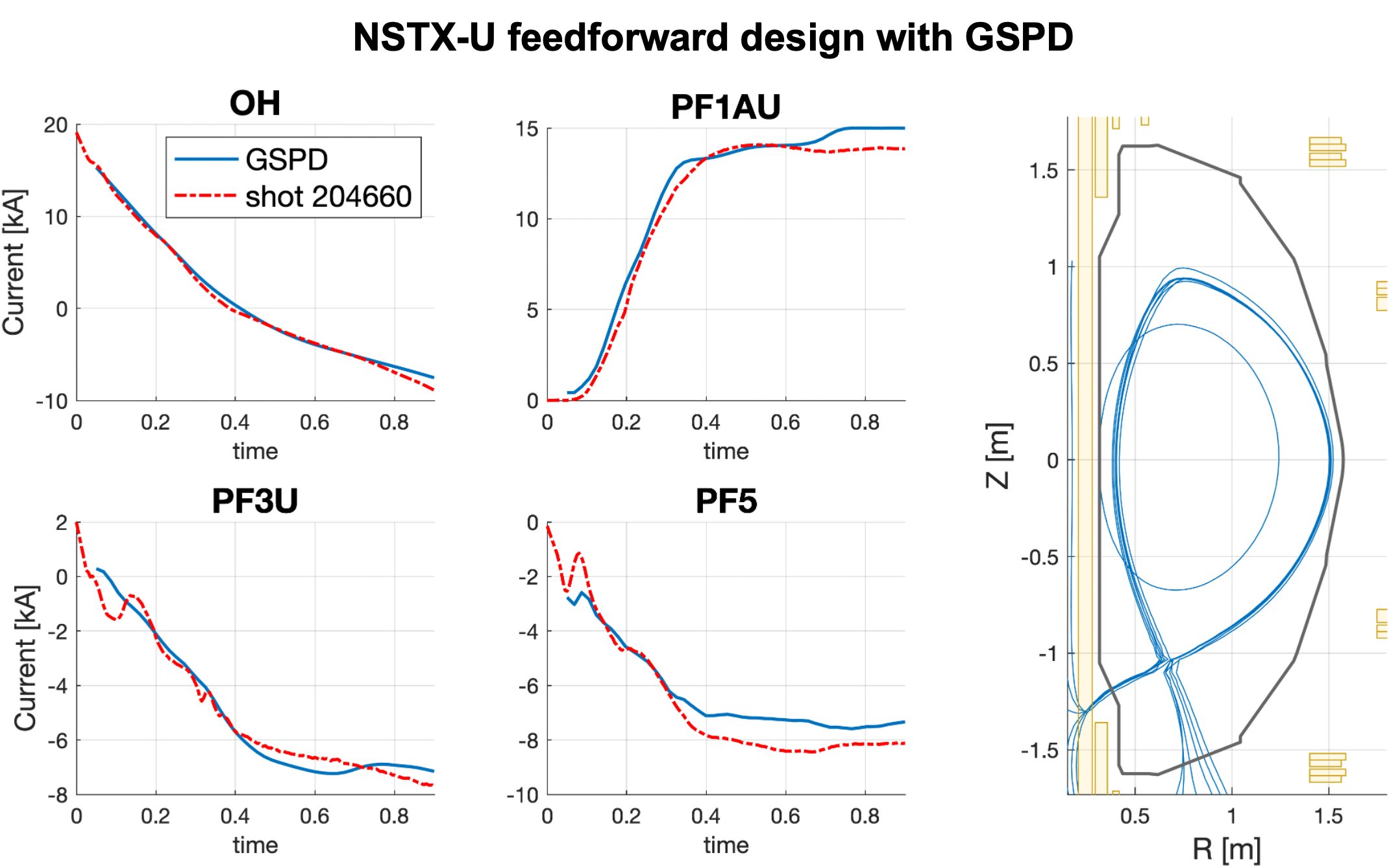}}
    \end{center}
    \caption{A GSPD sequence designed to match the shapes of NSTX-U 204660. This captures several important features such as the current ramps to sustain plasma current, and that the OH ramp does not supply a uniform flux distribution so PF3 continues to ramp throughout the pulse. As a result, the main boundary stays constant but the strike point moves throughout the discharge. }
    \label{fig:nstxu_gspd}
\end{figure}

An example of GSPD for a NSTX-U shot design is shown in \cref{fig:nstxu_gspd}, which attempted to match the shaping trajectory of shot 204660. The designed equilibrium sequence shows good agreement with the experimental currents, and captures various important physical effects. For example, the OH coil is required to ramp continually throughout the entire shot to hold the plasma current constant. Additionally, the PF3 current is also ramping throughout the entire shot, to compensate for shaping changes introduced by the OH coil ramp. The right side of \cref{fig:nstxu_gspd} shows the plasma boundary at various times in the sequence. One notable effect, which is also true of NSTX-U experiments, is that the strike point changes throughout the shot even while attempting to hold the shape constant. This is from the fact that the OH coil cannot supply a completely uniform flux for current drive, so that the strike point moves with the the OH ramp, even as the other coils adjust to keep the outer boundary constant. 

To run GSPD requires performing the following steps:

\begin{enumerate}
    \item Defining an initial equilibrium and currents. The user must specify the initial condition including vessel currents if any. This could be, for example, an equilibrium from right after plasma breakdown. 

    \item Define the sequence of target shapes, similar to what is needed for controlling a pulse.
    
    \item Define trajectories for several plasma scalars, the plasma resistance $R_p(t)$, stored thermal energy $W(t)$, internal inductance $l_i(t)$, and target plasma current $I_p(t)$. $W$, $l_i$, and $I_p$ are used to specify the profiles in the GS equation by scaling coefficients of profile basis functions. $R_p$ and $I_p$ are used to compute the flux consumption required, and Ip is also a target for the trajectory optimization. 

    \item Define weight and settings. There are several parameters and settings required for tuning the system, such as the number of equilibria to solve for, and specifying optimization weights for different parameters.
\end{enumerate}

\begin{figure}[H] 
    \begin{center}
        \includegraphics[width=13cm]{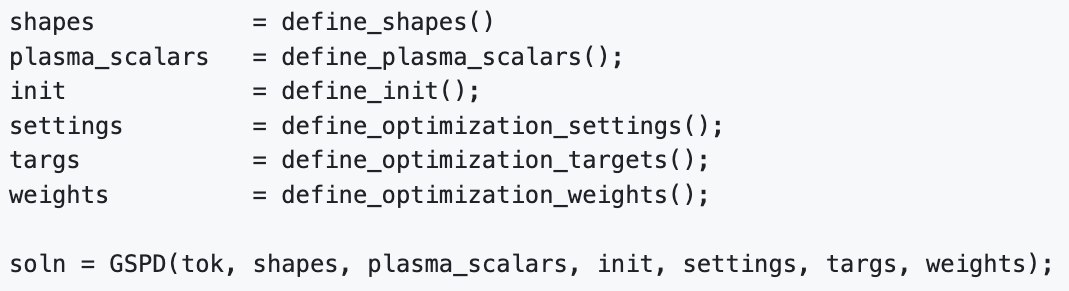}
    \end{center}
    \caption{Code snippet illustrating the main inputs and high-level function call. The code is available at \blue{github.com/PlasmaControl/GSPD.}}
    \label{fig:}
\end{figure}

\section{GSPD algorithm}

This section describes the algorithm in mathematical detail. Each equilibrium must satisfy the Grad-Shafranov force balance equation:

\begin{equation}\label{A1eq:GS}
    \begin{split}
        \Delta^* \psi &= -\mu_0 R J_\phi \\
        J_\phi &=  RP'(\psi) + \frac{FF'(\psi)}{\mu_0 R}
    \end{split}    
\end{equation}

The circuit dynamics describing the system are:

\begin{equation}\label{A1eq:circuit}
    v_i = R_i I_i + M_{ij} \dot I_j + \dot \Phi_i^{pla} 
\end{equation}

Here, $v$ is the power supply voltage ($v=0$ if $i$ corresponds to a a vessel element), $R$ is the resistance, $I$ is the current in each conductor, $M_{ij}$ is the mutual inductance between conductors, and $\dot \Phi_i^{pla}$ is the flux at the conductor due to motion of the plasma. 

In the shape control literature, this last term is often obtained via `linearizing an equilibrium' with the representation:

\begin{equation}
    \dot \Phi_i^{pla} = \pdv{ \Phi_i^{pla}}{I_j} \dot I_j := X_{ij} \dot I_j
\end{equation}

However, for the purposes here we use the representation:

\begin{equation}\label{A1eq:phi_pla}
    \dot \Phi_i^{pla} = M_{ig} \dot I_g
\end{equation}

where $M_{ig}$ is the mutual inductance between conductor $i$ and each grid location 
 $g$, and $I_g$ is the plasma current in grid cell $g$. (Note the connection to the Grad-Shafranov equation: $I_g$ is the plasma current distribution corresponding to the current density $J_\phi$). The difference between these two representations is that the former needs to use a model to perform the linearization but is able to represent the plasma flux in terms of the conductor current evolution; the latter does not perform any linearization but it requires knowledge of the evolution of the plasma current distribution. 

 The conductor evolution is written for both coil and vessel elements. Combining \cref{A1eq:circuit,A1eq:phi_pla} and being explicit about coils or vessel elements, we arrive at:

\begin{equation}\label{A1eq:cond_dynamics}
    \begin{bmatrix} V_c \\ 0 \end{bmatrix} = R_{cv} \begin{bmatrix} I_c \\ I_v \end{bmatrix} + M_{cv,cv} \begin{bmatrix} \dot I_c \\ \dot I_v \end{bmatrix} + M_{cv,g} \dot I_g 
\end{equation}

Lastly, using the Poynting flux method described in \cite{Ejima1982}, the total plasma current evolution is written in terms of the plasma surface voltage and flux at the boundary.

\begin{equation}\label{A1eq:psibry_dynamics}
    V_p = -\dot \psi_{bry} = R_p I_p + \frac{1}{I_p}  \dv{}{t} \left(\frac{1}{2}L_I I_p^2 \right) 
\end{equation}

The output of the pulse design algorithm is a set of equilibria that satisfy the governing \cref{A1eq:GS,A1eq:cond_dynamics,A1eq:psibry_dynamics}. 

The user inputs to the algorithm are targets for the plasma shape, total plasma current, expected plasma resistance, and the starting conductor currents. Also needed is pressure/current profile information to constrain the Grad-Shafranov solution. For this implementation we use a set of profile basis functions that are scaled to match user input of total stored thermal energy $W^{th}$, plasma current $I_p$, and normalized internal inductance $l_i$. 

The algorithm performs the following steps, which will be described individually below:

\begin{enumerate}
    
    \item \textbf{Initialization:} based on the target shape and $I_p$, the algorithm obtains a rough estimate of the plasma current distribution at each time step. 
    
    \item \textbf{Calculate target boundary flux:} use \cref{A1eq:psibry_dynamics} to obtain the target $\psi_{bry}$ at each time step. 
    
    \item \textbf{Optimize conductor evolution:} rewrite the conductor dynamics \cref{A1eq:cond_dynamics} in terms of a quadratic optimization problem. Solve for the conductor evolution that minimizes all shaping errors. 

    \item \textbf{Grad-Shafranov iteration:} The previous step also outputs the flux $\psi$ at each time for the current iteration. Perform Picard iteration for the plasma current distribution ($n$ indicating the iteration index):
    
    \begin{equation}
        J_\phi^{n+1} =  RP'(\psi^n) + \frac{FF'(\psi^n)}{\mu_0 R}
    \end{equation}

    \item \textbf{Repeat from step 2 until convergence.}
    
\end{enumerate}

Each step of the algorithm will now be explained individually

\subsectionbf{Initialization}

The purpose of this step is to obtain a rough estimate of the plasma current distribution. We use the target shape boundary and find the geometric centroid. Then each point on the grid is written in terms of a scaled distance x with x=0 at the centroid and x=1 at the boundary. The current is estimated with parabolic distribution:

\begin{equation}
    J_\phi = \hat J (1-x^a)^b
\end{equation}

where $a$ and $b$ are constants and $\hat J$ is a constant scaled to match the target Ip. 

\subsectionbf{Calculate target boundary flux}

The plasma internal inductance $L_I$ can be measured from the current distribution. With initial condition $\psi_{bry}(t=0)$ obtained from the starting equilibrium, \cref{A1eq:psibry_dynamics} can be integrated to find the $\psi_{bry}(t)$ that provides the surface voltage to drive the target Ip. 

\subsectionbf{Optimize conductor evolution}

This step updates the coil currents and applied flux, and is done by casting the update into the form of a quadratic optimization problem similar to model predictive control. 

The first step is to write \cref{A1eq:cond_dynamics} into a state space form:

\begin{equation}
    \dot x = A x + B u + w 
\end{equation}

where we have used the following substitutions

\begin{equation}    
\begin{split}   
    x &= [I_c^T \;\; I_v^T]^T \\
    u &= V_c \\
    A &= -M_{cv,cv}^{-1}R_{cv} \\
    B &= M_{cv,cv}^{-1} \\
    w &= -M_{cv,cv}^{-1}M_{cv,g} \dot I_g
\end{split}    
\end{equation}

Note that $w$ represents the influence of the plasma current on the conductors and can be calculated directly given the plasma current distribution at each time. Except for $w$ which depends on the Grad-Shafranov solutions and is updated each iteration, the remaining dynamic terms are linear (i.e. $A$ and $B$ do not change). This dynamics model is converted to discrete-time using the zero order hold method. For conciseness, we abuse the notation by using the same labels, making note that A, B, and w from hereon refer to their discrete time versions. 

\begin{equation}\label{A1eq:discrete_dynamics}
    x_{k+1} = A x_k + B u_k + w_k
\end{equation}

The shaping targets include parameters such as the flux at each of the control points, flux at target boundary location, and field at the target x-points. For any output $y$ that is a linear function $f$ of the grid flux distribution $\psi_g$ (which is true for flux and field measurements) then the output can be represented:

\begin{equation}
    \begin{split}
        y &= f(\psi_g) \\
          &= f(\psi_g^{app}) + f(\psi_g^{pla}) \\        
          &= f(M_{g,cv}I_{cv}) + f(\psi_g^{pla}) \\
          &:= f(M_{g,cv}I_{cv}) + y^{pla}
    \end{split}
\end{equation}

In other words, the output is separated into a part that depends on the coil and vessel currents and part that depends on the plasma flux distribution. Moreover, for flux and field measurements the function $f(\cdot)$ is a linear mapping (e.g. $f(x) = Ax$ for some A) so that we can write

\begin{equation}
    y = CI_{cv} + y^{pla} \\
\end{equation}

or using the state-space notation

\begin{equation}\label{A1eq:output}
    y_k = C x_k + y_k^{pla}
\end{equation}

The dynamics \cref{A1eq:discrete_dynamics} and output \cref{A1eq:output} form the basis of the prediction model for the optimization. In both equations we have intentionally separated the linear and nonlinear portions. The nonlinear terms $w_k$ and $y_k^{pla}$ are measured from the plasma current distribution and updated each iteration. The linear terms are updated during this optimization step by solving an MPC-like optimization problem. 

We use as a cost function for the optimizer

\begin{equation}
    J = \sum_{k=1}^{N} u_k^T R_u u_k + \Delta u_k^T R_{\Delta u} u_k + e_{k+1}^T Q_e e_{k+1} + \Delta \hat e^T \hat Q_{\Delta e} \Delta \hat e
\end{equation}

Where $e_k = r_k - y_k$ is the shape error between the reference target and actual, and $Q_e, R_u, R_{\Delta u}$ are user-defined weights on the shape errors, voltage inputs, and derivative of voltage inputs. 

The next few steps map the cost function into a standard quadratic program which can be solved by many available software packages such as MatLab's \textbf{quadprog}. We will transform the above cost function into the standard quadprog form 

\begin{equation}\label{A1eq:standardform}
    J = \hat u^T H \hat u + 2f^T \hat u
\end{equation}

To begin, we make the following definitions

\begin{equation}
    \begin{split}
        \hat u & := [u_1^T \; u_2^T \; ... \; u_{N}^T]^T \\
        \hat {\Delta u} & := [(u_1-u_0)^T \;\; (u_2-u_1)^T \; ... \; (u_N-u_{N-1})^T]^T \\
        \hat w & := [w_1^T \; w_2^T \; ... \; w_{N}^T]^T \\
        \hat x & := [x_2^T \; x_3^T \; ... \; x_{N+1}^T]^T \\        
        \hat y & := [y_2^T \; y_3^T \; ... \; y_{N+1}^T]^T \\        
        \hat r & := [r_2^T \; r_3^T \; ... \; r_{N+1}^T]^T \\        
        \hat e & := [e_2^T \; e_3^T \; ... \; e_{N+1}^T]^T \\       
        \hat Q_e &:= \textbf{blkdiag(} \underbrace{Q_e, Q_e \; ... \; Q_e}_{\times N} \textbf{)} \\
        \hat R_u &:= \textbf{blkdiag(} \underbrace{R_u, R_u \; ... \; R_u}_{\times N} \textbf{)} \\
        \hat R_{\Delta u} &:= \textbf{blkdiag(} \underbrace{R_{\Delta u}, R_{\Delta u} \; ... \; R_{\Delta u}}_{\times N} \textbf{)} \\
        \hat C &:= \textbf{blkdiag(} \underbrace{C, C \; ... \; C}_{\times N} \textbf{)}
    \end{split}
\end{equation}

Each term in the cost function will now be treated separately. The first term gives

\begin{equation}\label{A1eq:J1}
    J_1 = \sum_{k=0}^N u_k^T R_u u_k = \hat u^T \hat R \hat u
\end{equation}

For the second term we note that 

\begin{equation}
    \hat{\Delta u} = \underbrace{\begin{bmatrix}
        I \\
        -I & I \\
        & & \ddots \\
        & & -I & I \\
    \end{bmatrix}}_{:=S_u} \hat u - 
    \underbrace{\begin{bmatrix}
        u_0 \\ \\ \\ \\ 
    \end{bmatrix}}_{:=u_{prev}}
\end{equation}

so that the 2nd term in the cost function is:

\begin{equation}\label{A1eq:J2}
\begin{split}
    J_2 &= \sum_{k=0}^N \Delta u_k^T R_{\Delta u} \Delta u_k \\
    &= \hat u^T S_u^T \hat R_{\Delta_u} S_u \hat u - 2u_{prev}^T R_{\Delta_u} S_u \hat u
\end{split}    
\end{equation}

For the last term, we use the dynamics model \cref{A1eq:discrete_dynamics} to write

\begin{equation}
    \nonumber
    \begin{bmatrix} x_{2} \\ x_{3} \\ \vdots \\ x_{N+1} \end{bmatrix} 
    =
    \underbrace{\begin{bmatrix} A \\ A^2 \\ \vdots \\ A^N \end{bmatrix}}_{E} x_1
    +
    \underbrace{\begin{bmatrix}  B \\  AB & B \\ \vdots & \vdots & \ddots \\  A^{N-1}B &  A^{N-2}B & \hdots & B \end{bmatrix}}_{F}
    \begin{bmatrix} u_1 \\ u_{2} \\ \vdots \\ u_{N} \end{bmatrix}
    + \underbrace{\begin{bmatrix} I  \\  A & I \\ \vdots & \vdots & \ddots \\  A^{N-1} &  A^{N-2} & \hdots & I \end{bmatrix}}_{F_w}
    \begin{bmatrix} w_1 \\ w_2 \\ \vdots \\ w_{N} \end{bmatrix}    
\end{equation}

Or equivalently

\begin{equation}\label{pm}
    \hat x = Ex_1 + F \hat u + F_w \hat w
\end{equation}

Then using \cref{A1eq:output} the predicted errors are:

\begin{equation}
\begin{split}
    \hat e &= \hat r - \hat y \\
        &= \hat r - \hat y^{pla} - \hat C \hat x \\
        &= \hat r - \hat y^{pla} - \hat C (Ex_1 + F \hat u + F_w \hat w) \\
        &:= M \hat u + d 
\end{split}    
\end{equation}

In the last step we have defined appropriate variables $M$ and $d$ for compactness. Then the third term in the shaping cost function is

\begin{equation}\label{A1eq:J3}
\begin{split}
    J_3 &= \sum_{k=0}^N e_{k+1}^T Q_e e_{k+1} \\
    &= \hat u^T \left[ M^T \hat Q_e M \right ] \hat u + 2 \left[ d^T Q_e M \right ]\hat u
\end{split}    
\end{equation}

\begin{equation}
    \Delta \hat e = S_e \hat e - e_{prev}
\end{equation}

\begin{equation}
    e_k = r_k - y_k^{pla} - Cx_k
\end{equation}

\begin{equation}
\begin{split}
     J_4 &= \Delta \hat e^T \hat Q_{\Delta e} \Delta \hat e  \\ &=
     \left[ S_e(M \hat u + d) - e_{prev}  \right]^T \hat Q_{\Delta e} \left[ S_e(M \hat u + d) - e_{prev}  \right] \\
     &= \hat u^T \left [ M^T S_e^T \hat Q_{\Delta e} S_e M \right ] \hat u + 2 \left[ M^T S_e^T \hat Q_{\Delta e} (S_e d - e_{prev} )\right ] \hat u
\end{split}
\end{equation}

Each term (\cref{A1eq:J1,A1eq:J2,A1eq:J3} in the cost function has now been written in the standard quadratic program form \cref{A1eq:standardform}. We use a quadratic program solver to find the coil and vessel trajectories, also including any constraints on the power supplies and shaping. From the coil and vessel trajectories we also compute the applied flux on the grid at each time step via the mutual inductances, which will be used in the next stage. 

\begin{equation}
    \psi_g^{app} = M_{g,cv} I_{cv}
\end{equation}

\subsectionbf{Free-boundary Grad-Shafranov iteration}

In this stage of the solver, we perform a Picard iteration to update the plasma current density distribution according to the Grad-Shafranov equation:

\begin{equation}\label{A1eq:GS_iterate}
    J_\phi^{k+1} =  RP'(\psi^k) + \frac{FF'(\psi^k)}{\mu_0 R}
\end{equation}

\begin{figure}[H] 
    \begin{center}
        \includegraphics[width=8cm]{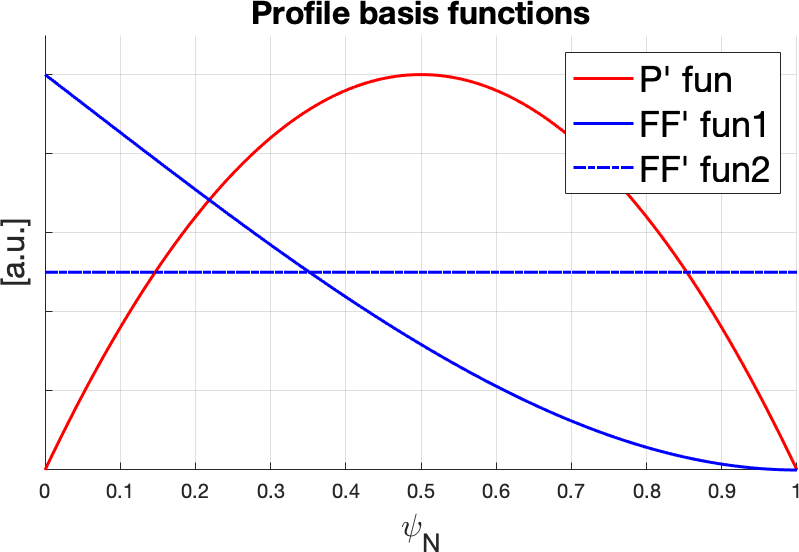}
    \end{center}
    \caption{Polynomial basis functions for Grad-Shafranov solution.}
    \label{A1fig:basis_funs}
\end{figure}

The flux distribution has just been updated with new externally applied flux from the conductor evolution. For this implementation, we use polynomial basis functions which are shown in \cref{A1fig:basis_funs} to describe the $P'$ and $FF'$ functions. The total $P'$ and $FF'$ is then the product of the basis functions (subscript $b$) and several coefficients which will be solved for. 

\begin{equation}
\begin{split}
    P'(\psi_N) &= c_p P'_{b} \\
    FF'(\psi_N) &= c_{f1} FF'_{b1} + c_{f2} FF'_{b2}
\end{split}
\end{equation}

This form with 3 coefficients is less complex than what is used in most transport models, although 3 coefficients has been shown to be accurate enough for performing shape control with simplified transport \cite{Welander2019}, and future work may expand this to allow more profile shaping capability. The basis functions are written in terms of the normalized flux which is normalized relative to the flux at the boundary and magnetic axis. 

\begin{equation}
    \psi_N := \frac{\psi - \psi_{bry}}{\psi_{mag} - \psi_{bry}}
\end{equation}

This introduces a scaling factor 
\begin{equation}
    P'(\psi) = \frac{P'(\psi_N)}{\psi_{mag} - \psi_{bry}}
\end{equation}

and similarly for the $FF'$ profile. 

We use a fieldline tracing algorithm to identify x-points or touch points of the plasma, identify the boundary-defining point, and trace the boundary. This also calculates $\psi_{mag}$ and $\psi_{bry}$. 

The profile coefficients are found by matching a target $I_p$, thermal energy $W^{th}$, and normalized internal inductance $l_i$.

For matching $I_p$ we have that 

\begin{equation}
    I_p = \int J_\phi dA
\end{equation}

Using the Grad-Shafranov equation this is written as

\begin{equation}\label{A1eq:Ip}
    I_p = \frac{1}{\psi_{mag}-\psi_{bry}} \left [ c_{p} \int RP'_b dA  + c_{f1} \int \frac{FF'_{b1}}{\mu_0 R} dA + c_{f2} \int \frac{FF'_{b2}}{\mu_0 R} dA  \right ]
\end{equation}

which depends linearly on the profile coefficients. For matching the thermal energy $W^{th}$ we use

\begin{equation}\label{A1eq:Wth}
    W^{th} = \int 3 \pi R P(\psi_N) dA \\   
\end{equation}

where the pressure is found from integrating the $P'$ basis function. 

\begin{equation}
    P(\psi_N) = c_{p} \int_0^{\psi_N} P'_b d \hat \psi_N
\end{equation}

Unlike the other two parameters, internal inductance $l_i$ cannot be written as a linear function of the profile coefficients. One solution is to calculate gradients and iterate to converge toward a target $l_i$. However, in this first release we instead enforce an alternative, simpler constraint on the $FF'$ coefficients:

\begin{equation}\label{A1eq:li}
    c_{f1} + \alpha c_{f2} = 0
\end{equation}

One of the $FF'$ basis functions is flat, and the other is peaked (see \cref{A1fig:basis_funs}) and so the ratio of the these two roughly determines how peaked the current profile is, which is also what the $l_i$ parameter indicates although it is defined differently. We first fit a few equilibria exactly and then measure the ratio of these to determine an approximate linear relationship for $\alpha(l_i)$. Future work will expand the capabilities for profile matching. 

The system of three target parameters ($I_p, W_{th}, l_i$) and three unknown coefficients ($c_p, c_{f1}, c_{f2})$ can now be solved, so that the full unnormalized profile functions are known. Then the Grad-Shafranov equation \cref{A1eq:GS_iterate} is updated which gives a new plasma current density distribution $J_\phi$. Note that the plasma current distribution $I_g$ is exactly the current density distribution $J_\phi$ scaled by the area of each grid cell. Then the flux from plasma sources on the grid is obtained: 

\begin{equation}
    \psi_g^{pla} = M_{gg} I_g 
\end{equation}

At this stage we have now updated the plasma current and plasma flux distributions, and the solver can proceed to the next iteration. This concludes the description of the algorithm.

\pagebreak

% ===================
% SECTION: CONCLUSION
% ===================
\section{Conclusion}

This work describes GSPD, a tool for time-dependent equilibrium design which includes dynamic effects such as induced vessel currents and plasma current drive. The algorithm is shown to reproduce experimental results and illustrates time-dependent phenomena of the equilibrium evolution, such as the shaping modifications throughout an NSTX-U discharge.

% =================
% ACKNOWLEDGEMENTS
% =================

\bibliography{bib}

\begin{thebibliography}{2}
\expandafter\ifx\csname natexlab\endcsname\relax\def\natexlab#1{#1}\fi
\providecommand{\url}[1]{\texttt{#1}}
\providecommand{\href}[2]{#2}
\providecommand{\path}[1]{#1}
\providecommand{\DOIprefix}{doi:}
\providecommand{\ArXivprefix}{arXiv:}
\providecommand{\URLprefix}{URL: }
\providecommand{\Pubmedprefix}{pmid:}
\providecommand{\doi}[1]{\href{http://dx.doi.org/#1}{\path{#1}}}
\providecommand{\Pubmed}[1]{\href{pmid:#1}{\path{#1}}}
\providecommand{\bibinfo}[2]{#2}
\ifx\xfnm\relax \def\xfnm[#1]{\unskip,\space#1}\fi
%Type = Article
\bibitem[{Ejima et~al.(1982)Ejima, Callis, Luxon, Stambaugh, Taylor, and
  Wesley}]{Ejima1982}
\bibinfo{author}{S.~Ejima}, \bibinfo{author}{R.~Callis},
  \bibinfo{author}{J.~Luxon}, \bibinfo{author}{R.~Stambaugh},
  \bibinfo{author}{T.~Taylor}, \bibinfo{author}{J.~Wesley},
\newblock \bibinfo{title}{Volt-second analysis and consumption in doublet {III}
  plasmas},
\newblock \bibinfo{journal}{Nuclear Fusion} \bibinfo{volume}{22}
  (\bibinfo{year}{1982}) \bibinfo{pages}{1313--1319}. \URLprefix
  \url{https://doi.org/10.1088/0029-5515/22/10/006}.
  \DOIprefix\doi{10.1088/0029-5515/22/10/006}.
%Type = Article
\bibitem[{Welander et~al.(2019)}]{Welander2019}
\bibinfo{author}{A.~Welander}, et~al.,
\newblock \bibinfo{title}{Closed-loop simulation with grad-shafranov
  equilibrium evolution for plasma control system development},
\newblock \bibinfo{journal}{Fusion Engineering and Design}
  \bibinfo{volume}{146} (\bibinfo{year}{2019}) \bibinfo{pages}{2361--2365}.
  \DOIprefix\doi{10.1016/j.fusengdes.2019.03.191}.

\end{thebibliography}

\end{document}